\documentclass[pre,superscriptaddress,showkeys]{revtex4}
\usepackage{graphicx}
\usepackage{amssymb}
\usepackage{url}
\usepackage{array}
\usepackage{epsfig}
\usepackage{url,hyperref}

\begin{document}

\title{Efficient data compression from statistical physics of codes over  finite fields}

\author{A Braunstein}
\affiliation {Dipartimento di Fisica and Center for Computational Sciences, Politecnico di Torino, Corso Duca degli Abruzzi 24, 10129 Torino, Italy}
\affiliation {HuGeF, Via Nizza 52, 10126 Torino, Italy}
\affiliation{Collegio Carlo Alberto, Via Real Collegio 30, 10024 Moncalieri, Italy}
\author{F Kayhan}
\affiliation{Dipartimento di Elettronica, Politecnico di Torino, Corso Duca degli Abruzzi 24, 10129 Torino, Italy}
\author{R Zecchina}
\affiliation {Dipartimento di Fisica and Center for Computational Sciences, Politecnico di Torino, Corso Duca degli Abruzzi 24, 10129 Torino, Italy}
\affiliation {HuGeF, Via Nizza 52, 10126 Torino, Italy}
\affiliation{Collegio Carlo Alberto, Via Real Collegio 30, 10024 Moncalieri, Italy}

\email{{alfredo.braunstein, farbod.kayhan, riccardo.zecchina}@polito.it}

\newcommand{\C}{\mathcal{C}}
\newcommand{\bfx}{\mathbf{x}}
\newcommand{\bfc}{\mathbf{c}}
\newcommand{\bfd}{\mathbf{d}}
\newcommand{\bfu}{\mathbf{u}}
\newcommand{\bfs}{\mathbf{s}}
\newcommand{\bfy}{\mathbf{y}}
\newcommand{\bfmu}{\mathbf{\mu}}

\newcommand{\bin}[2]{
    \left (
        \begin{array}{@{}c@{}}
        #1 \\ #2
        \end{array}
    \right )
}

\begin{abstract}
In this paper we discuss a novel data compression  technique  for binary
symmetric sources based on the cavity method over GF($q$), the Galois Field of
order $q$. We present a scheme of low complexity and  near optimal empirical
performance.

The compression step is based on a reduction of a  sparse low density
parity check codes over GF($q$) and is done through the so called 
reinforced belief-propagation equations.  These reduced codes appear to have a
non-trivial geometrical modification of the space of codewords which makes such
compression computationally feasible.
The computational complexity is $\mathcal{O}(d\cdot n\cdot q\cdot\log_{2} q)$
per iteration, where $d$ is the average degree of the check nodes and $n$ is
the number of bits. 

For our code ensemble, decompression can be done in a time linear in the code's
length by a simple leaf-removal algorithm.

\end{abstract}


\maketitle

\section{Introduction}
\label{sec:INTR} 

The relation between information theory and statistical mechanics of disordered
systems has been long stablished \cite{Sourlas,Nishimori}.  Since then, various
techniques from statistical physics of disordered systems have been used not
only to assess the theoretical bounds of the achievable performance but also to
provide practical encoding/decoding methods for lossy data compression.  In
particular, both cavity method and replica symmetry breaking techniques have
been used to demonstrate the Shannon results and assess the performance of
codes defined on sparse factor graphs \cite{Kaba1,Mimura,Murayama,Murayama2}. 


In this paper we address the classical problem of finding an efficient  lossy
compression scheme for a generic binary symmetric source. This objective is
reached by exploiting some unexpected  features of  the  cavity method  when
applied to graphical codes defined over   a finite field algebra of high order.

Given any realization $\bfy \in \{ 0,1 \}^n $ of a symmetric Bernoulli process    $\mathbf{Y}$, the goal is to compress
$\bfy$ by mapping it to a shorter binary vector such that an approximate reconstruction of $\bfy$ is possible within a given
fidelity criterion. More precisely, suppose $\bfy$ is mapped to the binary vector $\bfx \in \{ 0,1 \}^k$ with $k<n$ and $\hat{\bfy}$ is
the reconstructed source sequence. The quantity $R =k/n$ is called the compression rate. The fidelity or distortion is measured by
the Hamming distance $d_H (\bfy,\hat{\bfy}) = (1/n) \sum_{i=1}^{n} | y_i - \hat{y}_i|$. The goal is to minimize the average Hamming 
distortion $D = \mathbb{E} [d_H  (\mathbf{Y},\hat{\mathbf{Y}})]$ for any given rate. The asymptotic limit, known as the rate-distortion 
function, is given by $R(D) = 1 - H(D)$ for any $D \in [0,0.5]$ where $H(D) = -D\log_2 D - (1-D) \log_2 (1-D)$ is the binary entropy function.

Our approach in this paper is based on Low-Density Parity-Check (LDPC)
codes. Let $\mathcal{C}$ be a LDPC code with $k \times n$ generator
matrix $\mathbf{G}$ and $m \times n$ parity check matrix $\mathbf{H}$.
Encoding in lossy compression can be implemented like decoding in
error correction. Given a source sequence $\bfy$, we look for a
codeword $\hat{\bfy} \in \C$ such that $d_H(\bfy,\hat{\bfy})$ is
minimized. The compressed sequence $\bfx$ is obtained as the $k$
information bits that satisfy $\hat{\bfy} = \mathbf{G}^T \bfx$.

Even though LDPC codes have been successfully used for various types of
lossless data compression schemes \cite{CSV}, and also the existence of
ensembles that asymptotically achieve the Shannon's bound for binary symmetric
sources has been proved \cite{LDPC-Quantizer}, they have not been fully
explored for lossy data compression. It is partially due to the long standing
problem of finding a practical source-coding algorithm for LDPC codes, and
partially because Low-Density Generator Matrix (LDGM) codes, as dual of LDPC
codes, seemed to be more adapted for source coding and received more attention
in the few past years.

In \cite{MartinianYedidia}, Martinian and Yedidia show that quantizing
a ternary memoryless source coding with erasures is dual of the
transmission problem over a binary erasure channel.  They also prove
that LDGM codes, as dual of LDPC codes, combined with a modified
Belief Propagation (BP) algorithm can saturate the corresponding
rate-distortion bound. Following their pioneering work, LDGM codes
have been extensively studied for lossy compression by several
researchers \cite{lowerboundLDGM,F&F,LDGM-Urbank-LB,LDPC-LDGM,M-W2,WainWright}. 
In a series of parallel works, several researches
have used techniques from statistical physics to provide non-rigorous
analysis of LDGM codes \cite{Ciliberti,Kaba1,Murayama}. However, LDGM
codes seem to perform well only for rates smaller than 0.5. As we 
will see, our proposed LDPC codes perform very near to the 
rate distortion bound for rates larger than 0.5. For smaller rates 
the loss in performance can be compensated by increasing the 
complexity (number of iterations) of our coding scheme. 

In terms of practical algorithms, lossy compression is still an active research
topic. In particular, an asymptotically optimal low complexity compressor with
near optimal empirical performance has not been found yet.  Almost all
suggested algorithms have been based on some kind of decimation of
belief/survey propagation which suffers a computational complexity of
$\mathcal{O}(n^2)$ \cite{Ciliberti}, \cite{F&F} and \cite{WainWright}. One
exception is the algorithm proposed by Murayama \cite{Murayama}. When the
generator matrix is ultra sparse (US), the algorithm was empirically shown to
perform very near to the associated capacity needing $\mathcal{O}(n)$
computations. A generalized form of this algorithm, called Reinforced Belief
Propagation (RBP)\cite{BZ-prl} , was used in a dual setting
\cite{RBP-Blackwell}, for ultra sparse LDPC codes (US-LDPC) over GF($2$) for
lossy compression \cite{KayhanDP}. The main drawback in both cases is the
non-optimality of ultra sparse structures over GF($2$)
\cite{lowerboundLDGM,LDGM-Urbank-LB,Murayama}.  As we will see, this problem
can be overcome by increasing the size of the finite field.

Estimation of the weight enumerating function  show that randomly constructed
US-LDPC codes over GF($q$) nearly achieve the rate-distortion bound for $q \geq
64$.  Despite this, practical encoding for these codes is a hard task. The main
problem seems to stem from geometrical properties of the configuration space:
as the codes are good for channel coding, solutions are isolated and
well-separated. This characteristic is known to make the encoding problem difficult to
solve for iterative and local algorithms\cite{cluster, cluster2}. 
To improve this step we introduce the ensemble of \emph{b-reduced}
US-LDPC codes, which by eliminating a logarithmic number of constraints from
US-LDPC codes, just multiplies the number of codewords by a polynomial. This
change has a negligible effect in the rate, while having a large effect on the
performance of the scheme. Indeed, this modification not only improves the
convergence of the RBP algorithm on encoding, but also provides us with a
simple efficient decoding algorithm.

The rest of this paper is organized as follows. Section
\ref{sec:LDPCGF(q)} reviews the code ensemble which we use for lossy
compression. Section \ref{sec:RBPGF($q$)} describes the RBP algorithm
over GF($q$). We also discuss briefly the complexity and
implementation of the RBP algorithm. In section \ref{sec:ILC} we
describe iterative encoding and decoding for our ensemble and then
present the corresponding simulation results in section
\ref{sec:RESULT}. A brief discussion on further research is given in
Section \ref{sec:FR}.

\section{LDPC codes over GF($q$)}
\label{sec:LDPCGF(q)}


In this section we introduce the ultra sparse LDPC codes over
GF($q$). As we will see later, near Shannon's bound lossy compression is
possible using these codes and BP-like iterative algorithms.

\subsection{($\lambda,\rho$) Ensemble of GF($q$) LDPC codes}
We follow the methods and notations in \cite{irrLDPC} to construct
irregular bipartite factor graphs. What distinguishes GF($q$) LDPC
codes from their binary counterparts is that each edge ($i,j$) of
the factor graph has a label $h_{i,j} \in $ GF($q$) $ \setminus \{ 0
\}$. In other words, the non-zero elements of the parity-check
matrix of a GF($q$) LDPC codes are chosen from the non-zero elements
of the field GF($q$). Denoting the set of variable nodes adjacent to
a check node $j$ by $\mathcal{N}(j)$, a word $\bfc$ with components
in GF($q$) is a codeword if at each check node $j$ the equation
$\sum_{i \in \mathcal{N}(j)} h_{i,j} c_i = 0$ holds.

An ensemble of GF($q$) LDPC codes is characterized by two generating
polynomials
$\lambda(x) = \sum_{i=1}^{d_v}\lambda_i x^{i-1} \hspace{3 mm}
{\textrm{ and}} \hspace{3 mm} \rho(x) = \sum_{i=1}^{d_c}\rho_i
x^{i-1}$ where $\lambda_i$ ($\rho_i$) denotes the fraction of
edges incident on variable(check) nodes of degree $i$ and $d_v$
($d_c$) is maximum variable (check) node degree.

A ($\lambda,\rho$) GF($q$) LDPC code can be constructed from a ($\lambda,\rho$)
LDPC code by random independent and identically distributed selection of the
matrix coefficients with uniform probability from GF($q$)$\setminus \{ 0 \}$.
Note that this may not be an optimal way for selecting the coefficients. For
more details on code construction and coefficient selection we refer the
readers to \cite{NBLDPC1} and \cite{MacKay_Opt}. 


\subsection{Code Construction for Lossy Compression}
It is well known that the parity check matrix of a GF($q$) LDPC code,
optimized for binary input channels, is much sparser than the one of a
binary LDPC code with same parameters \cite{NBLDPC1,Davey-MacKay}.  In
particular, when $q \geq 2^6$, the best error rate results on binary
input channels is obtained with the lowest possible variable node
degrees, i.e., when almost all variable nodes have degree two.  Such
codes have been called \emph{ultra sparse} or \emph{cyclic} LDPC codes
in the literature. In the rest of this paper we call a LDPC code ultra
sparse (US-LDPC) if all variable nodes have degree two. We will mainly 
concentrate on codes in which the parity check's degree distribution 
is concentrated on at most two different degree values, for any given rate. 

Given a linear code $\C$ and an integer $b$, a $b$-reduction of $\C$
is the code obtained by randomly eliminating $b$ parity-check nodes of
$\C$. For reasons to be cleared in section \ref{sec:ILC}, we are
mainly interested in $b$-reduction of GF($q$) US-LDPC codes for small
values of $b$ ($1 \leq b \leq 5$).

GF($q$) US-LDPC codes have been extensively studied for transmission
over noisy channels \cite{HuElef}, \cite{DecodingNBLDPC},
\cite{Davey-MacKay}. The advantage of using such codes is twofold. On
the one hand, by moving to sufficiently large fields, it is possible
to obtain nearly capacity achieving codes.
On the other hand, the extreme sparseness of the factor graph is
well-suited for iterative message-passing decoding algorithms. Despite
the state of the art performance of moderate length GF($q$) US-LDPC
channel codes, they have been less studied for lossy compression. The
main reason being the lack of fast suboptimal algorithms. In the next
section we present RBP algorithm over GF($q$) and then show that practical
encoding for lossy compression is possible by using RBP as the encoding algorithm for the ensemble of $b$-reduced US-LDPC codes.

\section{Reinforced Belief Propagation Algorithm in GF($q$)}
\label{sec:RBPGF($q$)} In this section first we briefly review the
RBP equations over GF($q$) and then we discuss in some details the
complexity of the algorithm following Declercq and Fossorier
\cite{DecodingNBLDPC}.

\subsection{BP and RBP Equations}

The GF($q$) Belief Propagation (BP) algorithm is a straightforward
generalization of the binary case, where the messages are
q-dimensional vectors.

Let $\bfmu_{vf}^{\ell}$ denotes the message vector form variable node
$v$ to check node $f$ at the $\ell$th iteration. For each symbol $a
\in $GF($q$), the $a$th component of $\bfmu_{vf}^{\ell}$ is the
probability that variable $v$ takes the value $a$ and is denoted by
$\bfmu_{vf}^{\ell}(a)$. Similarly, $\bfmu_{fv}^{\ell}$ denotes the
message vector from check node $f$ to variable node $v$ at the
iteration $\ell$ and $\bfmu_{fv}^{\ell}(a)$ is its $a$th
component. Also let $\mathcal{N}(v)$ ($\mathcal{M}(f)$) denote the set
of check (variable) nodes adjacent to $v$ ($f$) in a given factor
graph.

Constants $\bfmu_{v}^{1}$ are initialized according to the prior information.
The BP updating rules can be expressed as follows:

{\bf Local Function to Variable:}
\begin{equation}\label{SPfunc2var}
    \bfmu_{fv}^{\ell}(a) \propto \sum_{{\textrm{Conf}}_{(v,f)}(a)}
    \;\; \prod_{ v' \in \mathcal{M}(f) \setminus \{v\}} \bfmu_{v'
    f}^{\ell}(a)
\end{equation}

{\bf Variable to Local Function:}
\begin{equation}\label{SPvar2func}
    \bfmu_{vf}^{\ell+1}(a) \propto \bfmu_{v}^{1}(a)
    \prod_{f' \in \mathcal{N}(v) \setminus
    \{f\}} \bfmu_{f'  v}^{\ell}(a)
\end{equation}

where ${\textrm{Conf}}_{(v,f)}(a)$ is the set of all
configurations of variables in $\mathcal{M}(f)$ which satisfy the
check node $f$ when the value of variable $v$ is fixed to $a$.
We define the marginal function of variable $v$ at iteration $\ell+1$
as

\begin{equation}\label{LocalField}
    \mathbf{g}_{v}^{\ell+1}(a) \propto \bfmu_{v}^{1}(a) \prod_{f \in \mathcal{N}(v)} \bfmu_{fv}^{\ell}(a).
\end{equation}

The algorithm converges after $t$ iterations if and only if for all
variables $v$ and all function nodes $f$
$$ \bfmu_{fv}^{t+1} = \bfmu_{fv}^{t}$$ up to some precision
$\epsilon$. A predefined maximum number of iterations $\ell_{\max}$
and the precision parameter $\epsilon$ are the input to the algorithm.

RBP is a generalization of BP in which the messages from variable
nodes to check nodes are modified as follows
\begin{equation}
\label{Eq:RBP} \bfmu_{vf}^{\ell+1}(a) \propto
\big(\mathbf{g}_{v}^{\ell}(a) \big)^{\gamma(\ell)}
     \bfmu_{v}^{1}(a) \prod_{f' \in \mathcal{N}(v) \setminus
    \{f\}} \bfmu_{f'  v}^{\ell}(a),
\end{equation}
where $\gamma(\ell):[0,1]\longrightarrow [0,1]$ is
a non-decreasing function and $\mathbf{g}_{v}^{\ell}$ is the marginal function of variable $v$
at iteration $\ell$. The marginals for RBP are defined as 

\begin{equation}\label{LocalField-RBP}
    \mathbf{g}_{v}^{\ell+1}(a) \propto 
    \big(\mathbf{g}_{v}^{\ell}(a) \big)^{\gamma(\ell)}
     \bfmu_{v}^{1}(a) \prod_{f \in \mathcal{N}(v)} \bfmu_{fv}^{\ell}(a).
\end{equation}

Intuitively, RBP equations can be thought as a sort of ``soft-decimation''
procedure.  Indeed, in a decimation procedure \cite{BMZ-rsa}, the BP equations
are iterated until convergence, and then an infinite external field with the
same sign of the local field is applied to one or more variables and the
process is repeated (until all variables receive an infinite field). In the
RBP procedure, every variable receives a finite external field which is
proportional to its own local field (the proportionality factor being $\gamma(\ell)$). Moreover, 
the two time-scales (convergence and external field update) are intermixed.

It is convenient to define $\gamma$ to be
\begin{equation}
\label{coolingfunction} \gamma(\ell) = 1 - \gamma_{0}\gamma_{1}^{\ell},
\end{equation}
where $\gamma_{0},\gamma_{1}$ are in $[0,1]$.

Note that when $\gamma_{1} = 1$, RBP is the same as the algorithm presented in
\cite{Murayama} for lossy data compression. In this case it is easy to show that the only fixed points of RBP are configurations that satisfy all the constraints.


\subsection{Efficient Implementation}

Ignoring the normalization factor in (\ref{SPvar2func}), to compute all
variable to check-node messages at a variable node of degree $d_v$ we need
$\mathcal{O} (q\cdot d_{v})$ computations. A naive implementation of GF($q$) BP
has computational complexity of $\mathcal{O}(d^{2}_{f}\cdot q^2)$ operations at
each check node of degree $d_{f}$. This high complexity is mainly due to the
sum in (\ref{SPfunc2var}), that can be interpreted as a discrete convolution of
probability density functions. Efficient implementations of function to
variable node messages based on Discrete Fourier Transform (DFT) have been
proposed by several authors, see for example
\cite{R&U,MacKay,NBLDPC1,DecodingNBLDPC} and the references within. The
procedure consists in using the identity 
$\bigodot_{v'\in \mathcal{M}(f)\setminus \{v\}}\bfmu_{v'f} = \mathcal{F}^{-1}\big(\prod_{ v' \in \mathcal{M}(f) \setminus \{v\}} \mathcal{F}\left(\bfmu_{v'f}\right)\big)$
where the $\bigodot$ symbol denotes convolution of functions over $GF(q)$, and the product on the right-hand side is the pointwise product of real-valued
functions.

Assuming $q=2^p$, the Fourier transform of each message $\bfmu_{v'f}$ needs
$\mathcal{O}(q\cdot p)$ computations and hence the total computational
complexity at check node $f$ can be reduced into $\mathcal{O}(d^{2}_{f}\cdot
q\cdot p)$. This complexity can be further reduced to $\mathcal{O}(d_{f}\cdot
q\cdot p)$ by using the fact that $\prod_{ v' \in \mathcal{M}(f) \setminus
\{v\}} \mathcal{F}\left(\bfmu_{vf}\right) = \prod_{ v' \in \mathcal{M}(f)}
\mathcal{F}\left(\bfmu_{v'f}\right)/\mathcal{F}\left(\bfmu_{vf}\right)$, or
alternatively by using the summation strategy described in \cite{BrMuPa} which
has the same complexity but is numerically more stable. Therefore, the total
number of computations per iteration is $\mathcal{O}(d\cdot q\cdot p\cdot n)$
where $d$ is the average check-node degree. A prototipe C++ implementation of 
these equations is provided in source form \cite{code}.

\section{Iterative Lossy Compression}
\label{sec:ILC}

In the following three subsections we first describe a simple method
for identifying information bits of a $b$-reduced US-LDPC code and
then present a near optimal scheme for iterative compression
(encoding) and linear decompression (decoding).

\subsection{Identifying a Set of Information Bits}

For $b$-reduced US-LDPC codes, one can use the \emph{leaf removal} (LR)
algorithm to find the information bits in linear time. In the rest
of this section we briefly review the LR algorithm and show that
1-reduction (removal of a sole check node) of a US-LDPC code
significantly changes the intrinsic structure of the factor graph of
the original code.

The main idea behind LR algorithm is that a variable on a leaf of a
factor graph can be fixed in such a way that the check node to which
it is connected is satisfied \cite{CoreRZ}. Given a factor graph, LR
starts from a leaf and removes it as well as the check node it is
connected to. LR continues this process until no leaf remains. The
residual sub-graph is called the \emph{core}. Note that the core is
independent of the order in which leaves (and hence the corresponding
check nodes) are removed from the factor graph. This implies that also
the number of steps needed to find the core does not depend on the
order on which leaves are chosen.

While US-LDPC codes have a complete core, i.e. there is no leaf in
their factor graph, the $b$-reduction of these codes have empty core.
Our simulations also indicate that even 1-reduction of a code largely
improves the encoding under RBP algorithm (see section
\ref{sec:RESULT}).  How RBP exploits this property is the subject of
ongoing research. 

As we have mentioned, the LR algorithm can be also used to find a set of
information bits of a given US-LDPC code. 
Let us examine the LR algorithm in more detail. At any step $t$ of LR algorithm, 
a leaf variable node $v_t$ attached to a factor node $f_t$ is selected. 
Denote by $F_t$ the remaining leaf variable nodes attached to check 
node $f_t$ ($F_t$ could be empty if there are no other leaves attached to it). 
Now we remove check node $f_t$ and leaf nodes in $F_t\cup \{v_t\}$, and repeat.
Under the hypothesis that the original graph was connected, this process is 
guarateed to finish at some time $T$ with the empty graph, as at each step except 
the last one, at least one leaf is created. 

It is easy to see that at each step, if we fix the values of all variables 
except those in $F_t\cup\{v_t\}$, then for each configuration of variables in $F_t$, 
the value of variable $v_t$ is uniquely determined. 
Therefore, $\cup_{t=1...T}F_t=\{w_1,\dots,w_{N-T}\}$ will form a set 
of information bits. Indeed, using the ordering of the variable 
indices $v_T,...,v_1,w_1,...,w_{N-T}$, the check matrix becomes 
upper triangular and each solution can be found by back substitution 
in linear time once information bits are fixed.

\subsection{Iterative Encoding} \label{subsec:Encoding} Suppose a code of rate
$R$ and a source sequence $\bfy$ is given. In order to find the codeword
$\hat{\bfy}$ that minimizes $d_H(\hat{\bfy},\bfy)$, we will employ the RBP
algorithm with a strong prior $\bfmu^1_{v}(a)=\exp(-L d_H(y_v,a))$ centered
around $\bfy$. The sequence of information bits of $\hat{\bfy}$ is the
compressed sequence and is denoted by $\bfx$.  In order to process the encoding
in GF($q$), we first need to map $\bfy$ into a sequence in GF($q$). This can be
simply done by grouping $p$ bits together and use the binary representation of
the symbols in GF($q$).

\subsection{Linear Decoding} \label{subsec:Decoding}

Given the sequence of information bits  $\bfx$, the goal of the decoder is to
find the corresponding codeword $\hat{\bfy}$.  This can be done by calculating
the $\mathbf{G}^T \bfx$ which in general needs $\mathcal{O}(n^2)$ computations.
One of the advantages of our scheme is that it allows for a linear complexity
iterative decoding. The decoding can be performed by iteratively fixing
variables following the inverse steps of the LR algorithm; at each step $t$
only one non-information bit is unknown (variable $v_t$) and its value can be determined from
the parity check $f_t$.  For a sparse parity-check matrix, the number of needed
operations is $\mathcal{O}(n)$. It is straightforward to show that a code has
empty core if and only if there exists a permutation of columns of the
corresponding parity-check matrix $\mathbf{H}$ such that $h_{ij} \neq 0$ for
$i=j$ and $h_{ij} = 0$ for all $i > j$. The decoding procedure is equivalent to
back-substitution on this permutated triangular matrix.


\section{Simulation Results} \label{sec:RESULT}
\subsection{Approximating the Weight Enumeration Function by BP}
\label{subsec:WEF}

Given an initial vector $\bfy$, and a probability distribution $P(\bfc)$
over all configurations, the $P$-average distance from $\bfy$ can be
computed by

\begin{equation}
D_{P}(\bfy) = \sum_{i}\sum_{c_i} P(c_i) d_{H}(c_i,y_i)\label{eq:avd}
\end{equation}

where $P(c_i)$ is the set of marginals of $P$. On the other hand, the
entropy of the distribution $P$ is defined by

\begin{equation}
S(P) = - \sum_{\bfc} P(\bfc) \log P(\bfc)\label{eq:S}.
\end{equation}

Even though it is a hard problem to calculate analytically both
marginals and $S(P)$ of a given code, one may approximate them using
messages of the BP algorithm at a fixed point \cite{Yedida}. Assuming
the normalized distance is asymptotically a self-averaging quantity
for our ensemble, $S(P)$ represents the logarithm of the number of
codeword at distance $D_{P}(\bfy)$ from $\bfy$. 
By applying a prior
distribution on codewords given by $\exp(-L d_H(\bfc,\bfy))$ one is able
to sample the sub-space of codewords at different distances from
$\bfy$.

\begin{figure}
\begin{centering}
    \includegraphics[angle=0,width=0.8\textwidth]{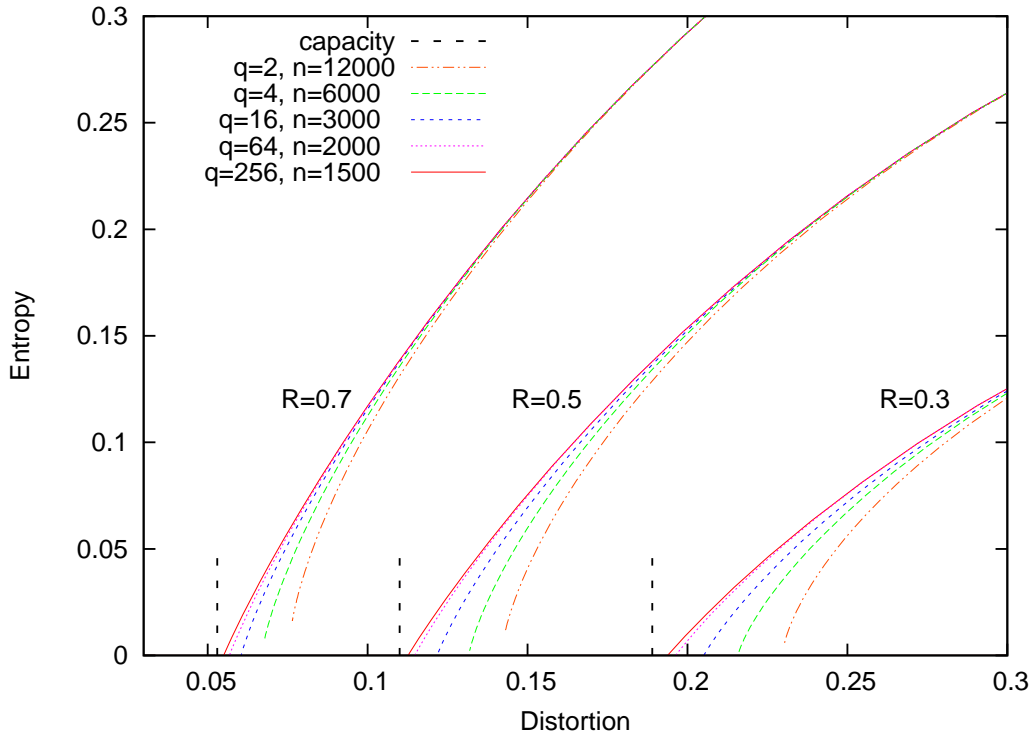}
    \caption {\label{Fig:WEF}(Color online) The approximate WEF of GF($q$) US-LDPC
    codes as a function of $q$ for a same blocklength in binary digits. }
\end{centering}
\end{figure}

Figure \ref{Fig:WEF} demonstrates the weight enumerator function (WEF) of
random GF(q) US-LDPC codes for rates 0.3, 0.5, and 0.7 and field orders 2, 4,
16, 64 and 256. The blocklength is normalized so that it corresponds to
$n=12000$ binary digits. 

Though BP is not exact over loopy graphs, we conjecture that the WEF
calculated for US-LDPC codes is asymptotically exact. This hypothesis
can be corroborated by comparing the plot in figure \ref{Fig:WEF} with
the simulation results we obtained by using RBP algorithm (figure
\ref{Fig:qPerformace}).

\subsection{Performance}
For the simplicity of the analysis, in all our simulations the parameter
$\gamma_1$ of RBP algorithm is fixed to one and therefore the function $\gamma$
is constant. We also fix the maximum number of iterations into $\ell_{max}
=300$. If RBP does not converge after 300 iterations, we simply restart RBP
with a new random scheduling.  The maximum number of trials allowed in our
simulations is $T_{max} =5$. The encoding performance depends on several
parameters such as $\gamma_0$,  $L$, the field order $q$, and the blocklength
$n$. In the following we first fix $n$, $q$ and $L$, in order to see how the
performance changes as a function of $\gamma_0$.  \subsubsection{Performance as
a Function of $\gamma_0$} \label{subsubsec:performancePEG} We will show that,
with this choice of
$\gamma(\ell)=1-\gamma_0$ there is a trade off, controlled by $\gamma_0$,
between three main aspects of the performance, namely: average distortion,
average number of iterations and average number of trials. The simulations in
this subsection are done for a 5-reduced GF(64) US-LDPC code with length
$n=1600$ and rate $R=0.33$. The factor graph is made by
\emph{Progressive-Edge-Growth} (PEG) construction \cite{HuElef}. The rate is
chosen purposefully from a region where our scheme has the weakest performance.
The Shannon's distortion bound for this rate is approximately $0.1754$. Note
that the non-monotonous behaviour of RBP as a function of $\gamma_0$ could be a
result of two concurrent phenomena: for small $\gamma_0$ the
reinforcement dynamics is too fast and may drive the system to non-codewords,
for large $\gamma_0$ the reinforcement contribution is small and the system does
not achieve polarization under the predefined iteration bound.
In the latter case, better performance may be achieved with $\gamma_1<1$ or
simply with a different choice of $\gamma(\ell)$.

\begin{figure}
\begin{centering}
    \includegraphics[angle=0,width=0.7\textwidth]{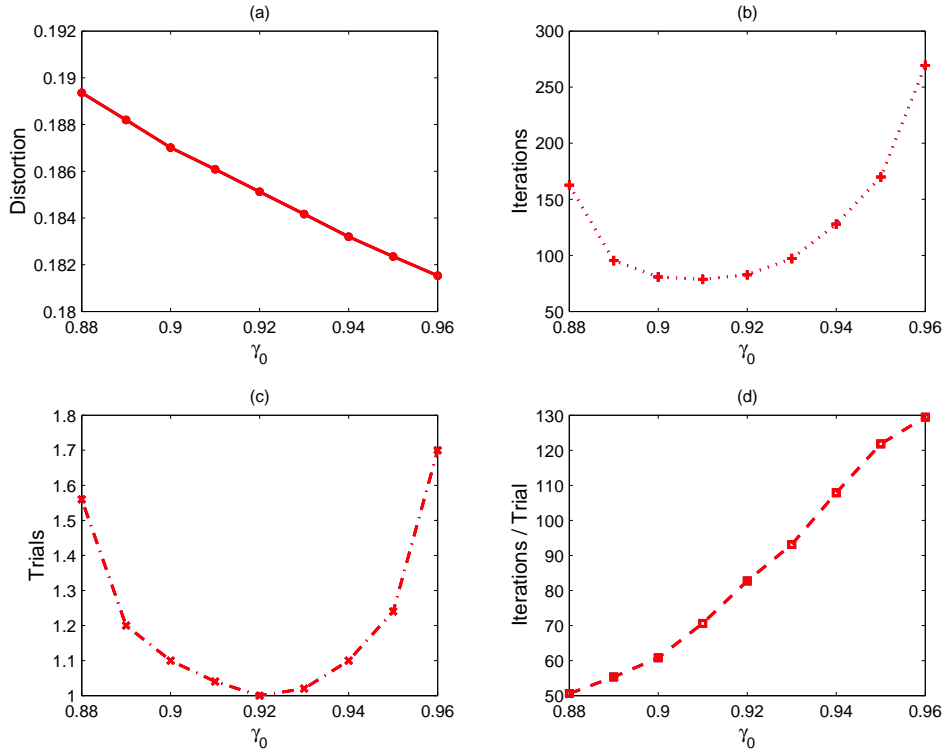}
    \caption {\label{Fig:gamma0} (Color online) Performance as a function of $\gamma_0$
    for a PEG graph with n=1600 and R=0.33. The averages are taken
    over 50 samples.(a) Average distortion as a function of $\gamma_0$.
    For $\gamma_0 > 0.96$ the RBP
    does not converge within 300 iterations. (b)The average number of
    iterations. (c)The average number of trials.  (d) The average number of iterations
    needed for each trial. Note that even though average number of iterations
    show a steep increase as a function of $\gamma_0$, the average number of
    iterations needed per trial increases only linearly.  }
\end{centering}
\end{figure}

In figure \ref{Fig:gamma0} we plot the performance as a function
of $\gamma_0$. For $\gamma_0 = 0.92$ we achieve a distortion of
$D=0.1851$ needing only 83 iterations in average and without any
need to restart RBP for 50 samples. By increasing $\gamma_0$ to 0.96,
one can achieve an average distortion of $0.1815$ which is only 0.15
dB away from the rate-distortion bound needing 270 iterations in average.
However, as it can be seen in figure \ref{Fig:gamma0}(d), the average 
number of iterations needed per trial increases only linearly as a function 
of $\gamma_0$.

\subsubsection{Performance as a function of $R$ and $q$}

\begin{figure}
\begin{centering}
    \includegraphics[angle=0,width=0.7\textwidth]{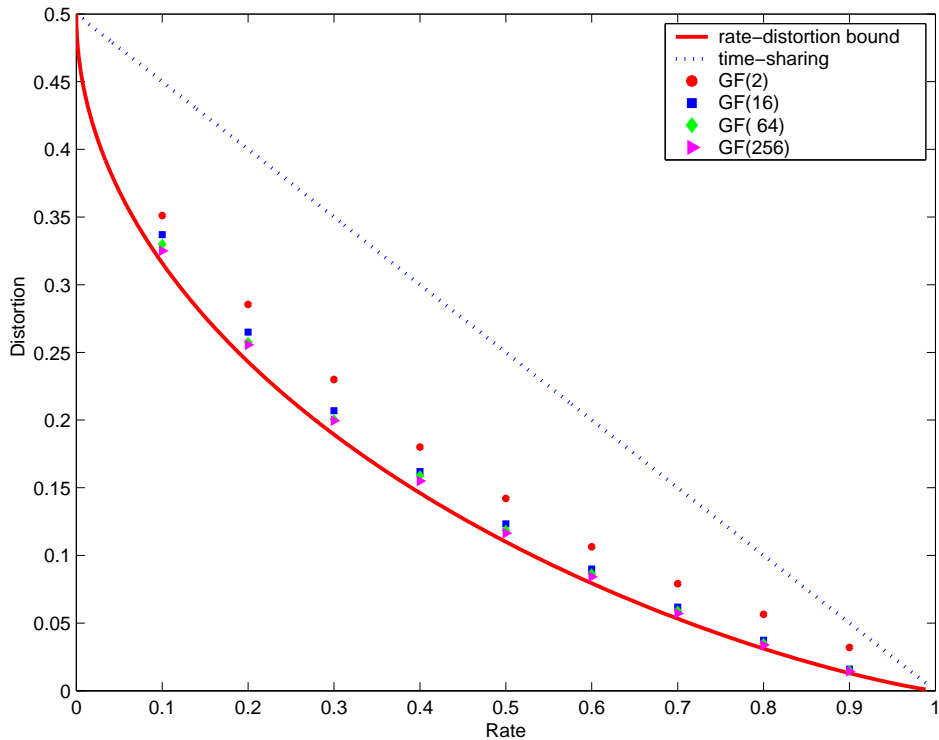}
    \caption {\label{Fig:qPerformace} (Color online) The rate-distortion performance
    of GF($q$) LDPC codes encoded with RBP algorithm for $q=2, 16, 64$ and $256$.
    The blocklength is
    12000 binary digits and each point is the average distortion over
    50 samples.}
\end{centering}
\end{figure}

Figure \ref{Fig:qPerformace} shows the distortion obtained by randomly
generated 5-reduced GF(q) US-LDPC codes for $q=2$, $q=16$, $q=64$ and
$q=256$. The block length is fixed to $n=12000$ binary digits. For each
given code with rate larger than or equal to 0.3, 
we choose $\gamma_0$ and $L$ so that the average number of
trials does not exceed 2 and the average number of iterations remains
less than 300. 
The optimized values of $\gamma_0$ and $L$ are found by
simulations and are reported in table \ref{tab:LGamma} for 
$q=256$. Under these two conditions, we report distortion
corresponding to best values of the two parameters averaged over 50
samples.  For codes with rates smaller than 0.3, one needs to 
allow for larger number of iterations and trials. 

As the data in table \ref{tab:LGamma} indicate, by increasing rate, both 
$L$ and $1 - \gamma_0$ increase. Larger values of $L$ impose stronger 
prior values, indicating that the initialized message distribution is 
more centered around $\bfy$. Note that in high rates, if $L$ is not 
chosen large enough, the loss in performance is substantial. On the other
hand, $\gamma_0$ regulates the reinforcement needed. Values very near 
to one for low rates indicates essentially the failure of reinforced 
strategy. This is not surprising, since in the absence of a codeword
near $\bfy$, forcing BP to find a solution is useless.

\begin{table}
\caption{The optimal values for $L$ and $\gamma_0$ obtained experimentally for $q = 256$.}
\begin{center}
\begin{tabular}{c |c c c c c c c c c c }
\hline
{\bf Rate} &{\bf 0.1}& {\bf 0.2} & {\bf 0.3 } & {\bf 0.4 }& {\bf 0.5}  & {\bf 0.6} & {\bf 0.7} & {\bf 0.8}&{\bf 0.9}\\
\\
 $L$  &1.1 &  1.3  & 1.5 & 1.7 & 1.9 & 2.3 & 2.4 & 2.8 & 3.8 \\
 \\
{\bf $\gamma_0$}&0.98 & 0.96  & 0.94 & 0.92 & 0.92 & 0.90 & 0.90 & 0.88 &0.88\\

\hline
\end{tabular}
\label{tab:LGamma}
\end{center}
\end{table}

\subsubsection{Reduction Effect on Performance of US-LDPC Codes}
\begin{figure}
\begin{centering}
    \includegraphics[angle=0,width=0.6\textwidth]{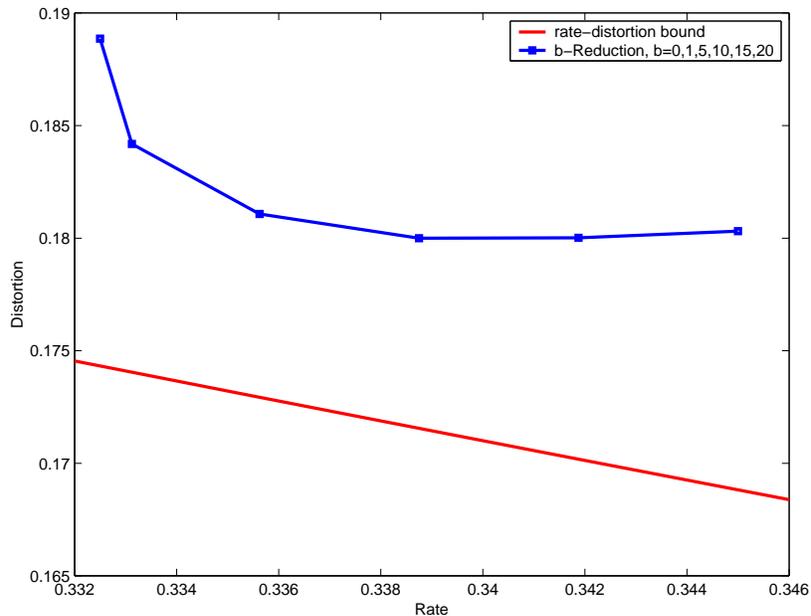}
    \caption {\label{Fig:breduction} (Color online) Performance as a function of $b$
    for a PEG graph with n=1600 and R=0.33 over GF(64). The averages are taken
    over 50 samples.}
\end{centering}
\end{figure}

As we have mentioned,  5-reduced LDPC codes have been used in our simulations.
The reduction improves both the convergence of RBP algorithm and 
the performance of the our scheme. In figure \ref{Fig:breduction} we show how 
the performance changes as a function of $b$. 
The simulations in this
subsection are done for a GF(64) US-LDPC code of length
$n=1600$ and rate $R=0.33$ with PEG constructed factor graph. 

\section{Discussion On Reduced Factor Graphs}
\label{sec:FR} Our results indicate that the scheme proposed in this
paper outperforms the existing methods for lossy compression by
low-density structures in both performance and complexity. The main
open problem is to understand and analyze the behaviour of RBP over
$b$-reduced US-LDPC codes.

We would like to add a few words to the role of $b$-reduction.  For simplicity,
let us concentrate on $q=2$, though the argument is general. First note that by
removing a parity check node from a code, the number of codewords is
doubled. This increment has an asymptotically negligible effect on the
compression rate since it only increases by $1/n$, while the robustness may
increase. More generally, it is possible to significantly alter the geometry of
the solution space while maintaining (asymptotically) the compression rate: for
instance, adding a path \{$\bfc=\bfd_1,\bfd_2,\dots,\bfd_k={\bf 0}$\} of new
codewords from each codeword $\bfc$ of a given code to the codeword ${\bf 0}$,
such that $d_H(\bfd_t,\bfd_{t+1})=1/n$ and $k\leq n$, multiplies the number of
codewords by at most $n$ and thus increases the rate by at most $\log n / n$
which is asymptotically negligible.  On the other hand, the codeword space
becomes ``star-shaped'' and thus connected on the hypercube geometry. Note
that such modified codes may be terrible for channel coding, as the
separation properties may have been severely worsened (e.g. the minimum
distance of the code becomes $1$). 

We think that a similar phenomenon could take place on $b$-reduced codes. On
the one hand, the asymptotic rate for source coding under the proposed scheme
is only increased by $b/n$ and the performance assuming MAP encoding can only
improve. On the other hand, we believe that the implied modification of the
geometry could ease the task of our iterative encoder. Indeed, it is well known
that large separation between solutions makes the problem very hard for
iterative and local algorithms \cite{cluster, cluster2}. In the following we
briefly explain some asymptotic implications of 1-reduction on weight
enumerating function of the US-LDPC code ensemble. 

\subsubsection{1-Reduced US-LDPC codes} As we have mentioned, cancelling a
single check node increases the the cardinality of the code by the factor $q$.
As we will see, for each codeword $\bfc$ of the original US-LDPC, there are
created $q-1$ new codewords which all have a distance $\mathcal{O}(\log n)$
from $\bfc$. In other words, a cluster of new codewords emerges for each
codeword $\bfc$.  In order to see this fact, let $v$ and $v'$ denote two
variables of degree one after removing the parity check $a$. With a probability
which approaches one, the checknode $a$ and both variables $v$ and $v'$ belong
to a loop of length $\mathcal{O} (\log n)$ of the original factor graph as $n
\to \infty$.  After removing $a$, this loop is broken and for any codeword of
the original factor graph one can obtain new codewords by assigning to $v$ any
value from the finite field and changing accordingly the values of all
variables in the broken loop. Note that this can be done because all variables
in the broken loop have degree two and $v'$ can be adjusted to satisfy the last
checknode in the path from $v$ to $v'$.

\section{Conclusions and perspectives}
\label{sec:Con}
Our main goal in this paper is to provide a low complexity coding scheme for
lossy data compression with near rate-distortion bound performance.  We propose
a practical iterative encoding/decoding scheme that exploits the geometrical
structure of the so called reduced ultra sparse low density parity check codes.
Our proposed algorithm for encoding can be considered as a soft decimation
strategy for belief propagation algorithm.  The complexity per iteration at the
iterative encoder depends linearly on both the length of the code and the order
of the field on which the code is defined.  The decoding algorithm is based on
leaf removal algorithm which has linear complexity on the proposed sparse
factor graphs.

We have investigated the behaviour of our scheme for various 
field orders and parameters of the proposed algorithm. In particular, 
we approximately calculate the weight enumerating function of US-LDPC codes 
as a function of field order using the BP algorithm. Our estimations show that 
US-LDPC codes over GF($q$) nearly achieve the rate-distortion bound for $q \geq
64$. Though BP is not exact over loopy graphs, we conjecture that the WEF
calculated for US-LDPC codes is asymptotically exact. This hypothesis
is corroborated by the simulation results we obtained by using RBP algorithm.

Our research can be expanded in several directions. For example, it is
interesting to study other ultra sparse ensembles sharing similar properties,
e.g. where just a certain fraction of variable nodes of degree one is allowed.
Several directions could be explored in order to obtain more efficient coding
schemes:  other choices of the reinforcement rate $\gamma(\ell)$, choices of
random codes and coefficient selection, and a $L\to\infty$ version of the
encoder along the lines of \cite{Declerc2} as it could allow much lower
computational complexity. Work is in progress in these direction.

\section*{Acknowledgment}

The authors wish to thank Guido Montorsi and Abolfazl Ramezanpour  for valuable suggestions and useful discussions. RZ acknowledges the ERC grant OPTINF 267915. The support from the EC grant STAMINA 265496 is also acknowledged AB and RZ.



\section*{References}
\begin{thebibliography}{10}

\bibitem{Sourlas} N.~Sourlas,~Nature {\bf 339},~693 (1989).

\bibitem{Nishimori} H.~Nishimori,  {\it Statistical Physics of Spin Glasses and Information Processing},  
(Oxford University Press,~UK,~2001).

\bibitem{Kaba1} T.~Hosaka and Y.~Kabashima,  Physica A 365,~113 (2006).

\bibitem{Mimura} K.~Mimura,  J. Phys. A: Math. Theor. {\bf 42},~135002 (2009).

\bibitem{Murayama} T.~Murayama,  Phys. Rev.~{\bf E 69},~035105R  (2004).
    
\bibitem{Murayama2} T.~Murayama  and M.~Okada,  J. Phys. A: Math. Gen. {\bf 36},  11123 (2003).
 
\bibitem{CSV}  G.~Caire, S~Shamai and S.~Verdu, {\it DIMACS Series in Discrete  Mathematics and 
Theoretical Computer Science}, (American Mathematical Society, 2004).

\bibitem{LDPC-Quantizer}  Y.~Matsunaga and  H.~Yamamoto, IEEE Trans. Information Theory {\bf 49}, 2225 (2003).

\bibitem{MartinianYedidia} E.~Martinian and J.~Yedidia, Proc. of Allerton Conference on 
Communication, Control and Computing,~131~(2003).

\bibitem{lowerboundLDGM}  A.~Dimakis, M.~Wainwright and K.~Ramchandran,  Proc. of the IEEE 
Inform. Theory Workshop,~650 (2007).

\bibitem{F&F} T.~Filler and J.~Fridrich, Proc. Allerton  Conference on Communication, Control and Comuting,~495~(2007).

\bibitem{LDGM-Urbank-LB} S.~Kudekar and R.~Urbanke, Proc. Int. Sympo. on Turbo Codes and Related Topics,~379~(2008).

\bibitem{LDPC-LDGM}  E.~Martinian and M.J.~Wainwright, IEEE Proc. Int. Symp. Info. Theory,~484 (2006).

\bibitem{M-W2} E.~Martinian and M.J.~Wainwright, IEEE Trans. on Information Theory {\bf 55},~1061 (2009).

\bibitem{WainWright}  M.J.~Wainwright, E.~Maneva and E.~Martinian, IEEE Trans. on Information Theory {\bf 56},~1351 (2010).

\bibitem{Ciliberti}   S.~Ciliberti, M.~Mezard and R.~Zecchina,  Phys. Rev. Lett. 95, 038701 (2005).

\bibitem{BZ-prl} A.~Braunstein, R.~Zecchina,  Phys. Rev. Lett.  96, 030201 (2006).

\bibitem{BMZ-rsa} A.~Braunstein, M.~Mézard, R.~Zecchina R. Random Struct. Algor. 27:201-26 (2005)
\bibitem{RBP-Blackwell}  A.~Braunstein, F.~Kayhan, G.~Montorsi and R.~Zecchina , IEEE Proc. Int. Symp. Info. Theory,~1891 (2007).

\bibitem{KayhanDP} F.~Kayhan and T.~Tanaka,  Proc. Int. Sympo. on Turbo Codes and Related Topics,~396 (2008).

\bibitem{cluster} L.~Dall'Asta, A.~Ramezanpour and R.~Zecchina, Phys. Rev. {\bf E 77}, 031118 (2008).

\bibitem{cluster2}  L.~Zdeborov{\'a} and F.~Krz{\'a}ka{\l}a, Phys. Rev. {\bf E 76}, 031131 (2007).

\bibitem{irrLDPC}  M.G.~Luby, M.~Mitzenmacher, M.A.~Shokrollahi and D.A~ Spielman,  IEEE Trans. Information Theory 47, 585 (2001).

\bibitem{NBLDPC1} A.~Bennatan and  D.~Burshtein,  IEEE Trans. Information Theory  52, 549 (2006).

\bibitem{MacKay_Opt} D.~MacKay, available at http://www.inference.phy.cam.ac.uk/mackay/CodesGallager.html,(2003).
    
\bibitem{Davey-MacKay} M.C.~Davey  and D.~MacKay,  IEEE Communication Letters  2 , 165 (1998).

\bibitem{HuElef}  X.Y.~Hu  and  E.~Eleftheriou, IEEE Proc. Conf. on Communications, 528 (2004).

\bibitem{DecodingNBLDPC}  D.~Declercq and M.~Fossorier,  IEEE Trans. Communication Theory 55, 633 (2007).

\bibitem{R&U} T.J.~Richardson and R.~Urbanke,  IEEE Trans. Information Theory 47, 599 (2001).

\bibitem{MacKay} D.J.C.~MacKay,  IEEE Trans. Information Theory 45, 399 (1999).

\bibitem{BrMuPa} A.~Braunstein, R.~Mulet and A.~Pagnani,   BMC Bioinformatics 9, 240 (2008).

\bibitem{CoreRZ}  M.~Mezard, F.~Ricci-Tersenghi  and R.~Zecchina, J. Stat. Phys. 111,  505 ( 2003).

\bibitem{Yedida} J.~Yedida, W.T.~Freeman and Y.~Weiss, Advances in Neural Information Processing Systems 13,~689 (2001).

\bibitem{Declerc2} D.~Declercq  and M~Fossorier, IEEE Proc. Int. Symp. Info. Theory, 464~(2005).




   
\bibitem{code}\url{http://www.polito.it/cmp/code/gfrbp} (2011).
\end{thebibliography}

\end{document}